# Fish aggregating devices drift like oceanographic drifters in the near-surface currents of the Atlantic and Indian Oceans

T. Imzilen[a], E. Chassot[b], J. Barde[c], H. Demarcq[a], A. Maufroy[a,d], L. Roa-Pascuali[a], J-F. Ternon[a], C. Lett[e,*]

[a] *MARBEC, IRD, Ifremer, Univ Montpellier, CNRS, Sète, France*
[b] *IRD, Seychelles Fishing Authority, PO BOX 449, Victoria, Seychelles*
[c] *Institut de Recherche pour le Développement (IRD), Indian Ocean Commission, Ébène, Mauritius*
[d] *ORTHONGEL, 11 bis rue des sardiniers, 29 900 Concarneau, France*
[e] *IRD, Sorbonne Université, UMMISCO, F-93143, Bondy, France*

**Abstract**

Knowledge of ocean surface dynamics is crucial for oceanographic and climate research. The satellite-tracked movements of hundreds of drifters deployed by research and voluntary observing vessels provide high-frequency and high-resolution information on near-surface currents around the globe. Consequently, they constitute a major component of the Global Ocean Observing System (GOOS). However, maintaining this array is costly and in some oceanic regions such as the tropics, spatio-temporal coverage is limited. Here, we demonstrate that the GPS-buoy equipped fish aggregating devices (FADs) used in tropical tuna fisheries to increase fishing success are also capable of providing comparable near-surface current information. We analyzed millions of position data collected between 2008 and 2014 from more than 15,000 FADs and 2,000 drifters, and combined this information with remotely-sensed near-surface current data to demonstrate that the surface velocity components of FADs and drifters are highly correlated in the Atlantic and Indian Oceans. While it was noted that the subsurface structures of FADs did slow them down relative to the drifters, particularly in the Atlantic Ocean, this bias was measurable and could be accounted for in future studies. Our findings show that the physical meteorological and oceanographic data collected by fishermen could provide an invaluable source of information to the GOOS. Furthermore, by forging closer collaborations with the fishing industry and ensuring their contributions to global ocean databases are properly acknowledged, there is significant scope to capture this data more effectively.

*Corresponding author
*Email address:* christophe.lett@ird.fr (C. Lett)

---

## Introduction

Oceans cover 70% of the Earth's surface and are much harder to observe than terrestrial systems (Richardson and Poloczanska, 2008). For centuries, mariners have been observing the states of oceans and the atmosphere by recording oceanographic and physical meteorological data near the ocean's surface (Woodruff et al., 1987). As early as the nineteenth century, international collaborative efforts were initiated to coordinate the collection and curation of ocean-atmosphere data from voluntary observing ships (VOS) and build large-scale marine data sets. Such data sets are now considered essential for oceanographic and climate studies (Woodruff et al., 1987; Kent et al., 2010; Freeman et al., 2017). From the 1970s, ocean data collection was revolutionized with the advent of satellite technology and the development of sensors that were capable of measuring a large range of oceanographic and atmospheric features (Martin, 2004).

Combining *in-situ* and remotely-sensed satellite observations has proven to be an essential step to improving our understanding of how ocean circulation affects climate at regional and global scales through the transport of water and heat received from the sun (Maximenko et al., 2009; Lee et al., 2010). Remotely-sensed measurements of sea surface temperature, altimetry and vector winds provide a synoptic view of ocean surface current patterns at consistent and regular spatial and temporal scales (Lagerloef et al., 1999; Sudre and Morrow, 2008; Dohan and Maximenko, 2010). At a finer scale, *in-situ* velocity measurements of near-surface currents are routinely collected by satellite-tracked drifters maintained by the Global Drifter Program (GDP), an operational component of the Global Ocean Observing System (GOOS) and the Global Climate Observing System (GCOS). This data provides a direct measurement of water properties and complements the satellite data by supplying information on high-frequency, small-scale oceanic processes (Niiler and Paduan, 1995; Reverdin et al., 2003; Lumpkin and Elipot, 2010). These drifters are floating devices that comprise a surface buoy equipped with a satellite transmitter and a subsurface sea anchor (Fig. 1). Since the 2010s, the GDP has maintained a global array of ∼1,200-1,500 drifters that have been deployed from VOS, research vessels and planes to cover the world's oceans (Joseph, 2013; Lumpkin and Johnson, 2013; Elipot et al., 2016). In addition to supporting oceanographic and climate research, the ocean circulation information acquired by these systems has been instrumental in supporting both military and civil applications, including search and rescue operations that use the data to improve their field of search predictions (Davidson et al., 2009). More recently, their role in tracking floating debris (Law et al., 2010; Cózar et al., 2014) has garnered attention as concerns about marine plastics pollution increase.

A knowledge of ocean dynamics is also key for fishermen who use it to both navigate and find fish resources. Monitoring surface water characteristics is essential in pelagic fisheries where the use of satellite remote-sensing has long been recognized as a fish harvesting aid (Simpson, 1992; Chassot et al., 2011). Modern fishing vessels are now equipped with a large range of sensors and electronic tools that constantly monitor the marine environment, enabling fishermen to identify the suitable habitats of target fish species (e.g. Torres-Irineo et al., 2014). Large-scale purse seiners are equipped with GPS and AIS positions systems, navigation compass, radars for both navigation and bird detection, sonars and lateral sounders for fish detection, current meters, wind sensors, and sea surface temperature (SST) thermometers. In addition, the vessels receive daily information on oceanographic features through commercial products derived from satellite imagery, i.e. meteorological and SST maps, sea-level anomaly data that allow identifying surface currents and temperature fronts as well as mesoscale features such as eddies and filaments, ocean-colour data, temperature data based on microwave imagery, and subsurface temperature maps (Saitoh et al., 2009). In tuna fisheries, the purse seine vessels that target fish schools have increasingly deployed satellite-tracked fish aggregating devices (FADs) over the last decade. Typically made of a bamboo raft equipped with floats to ensure buoyancy and a sea anchor built of old fishing nets (Fig. 1), these FADs attract tuna and increase fishery productivity (Fonteneau et al., 2013; Maufroy et al., 2017).

In recent years, the number of GPS-buoy equipped FADs used globally in this fishery has increased markedly. Currently, it is estimated that more than 100,000 FADs are drifting around the globe at any given time (Baske et al., 2012; Scott and Lopez, 2014). While the average lifespan of a FAD at sea (40 days; Maufroy et al., 2015) is shorter than a typical drifter (450 days; Lumpkin et al., 2012), there are many more in circulation, particularly in the tropical areas where the purse seine fleets operate. Consequently, it is likely that FADs could provide the GDP with complementary data, particularly in equatorial regions. Given that these areas are currently under sampled due to factors such as infrequent deployment of drifters and equatorial divergence (Lumpkin and Pazos, 2007), this increased FAD data coverage is especially important. As an illustrative case, a few FAD positions were used to complement the drifter data and ocean model outputs analyzed to locate the wreckage of the Air France flight that crashed in 2009 en route from Rio de Janeiro to Paris (Drévillon et al., 2013).

The overarching objective of this study is to test to what extent FADs deployed by fishermen are surrogates for GDP drifters, providing estimates of upper-ocean current velocities that are unbiased and of similar precision as those obtained from GDP drifters. To test this, we combined and analyzed large data sets from GDP drifters, a satellite-derived surface current product available from the Ocean Surface Currents Analyses Real-time (OSCAR) processing system and approximately 5 million FAD positions collected by French tuna fishing companies between 2008 and 2014 in the Atlantic and Indian Oceans.

## Material ans methods

To begin with, we directly compared the velocities of FAD and drifter pairs observed in close proximity over similar time periods. We then used the OSCAR currents as an indirect comparison point for both the FAD and drifter data. For the large biogeographical provinces of the Atlantic and Indian Oceans (Longhurst, 2007), we estimated the correlations between the OSCAR currents and the observed FAD and drifter velocities. We then compared FAD and drifter movements with short-term OSCAR current projections.

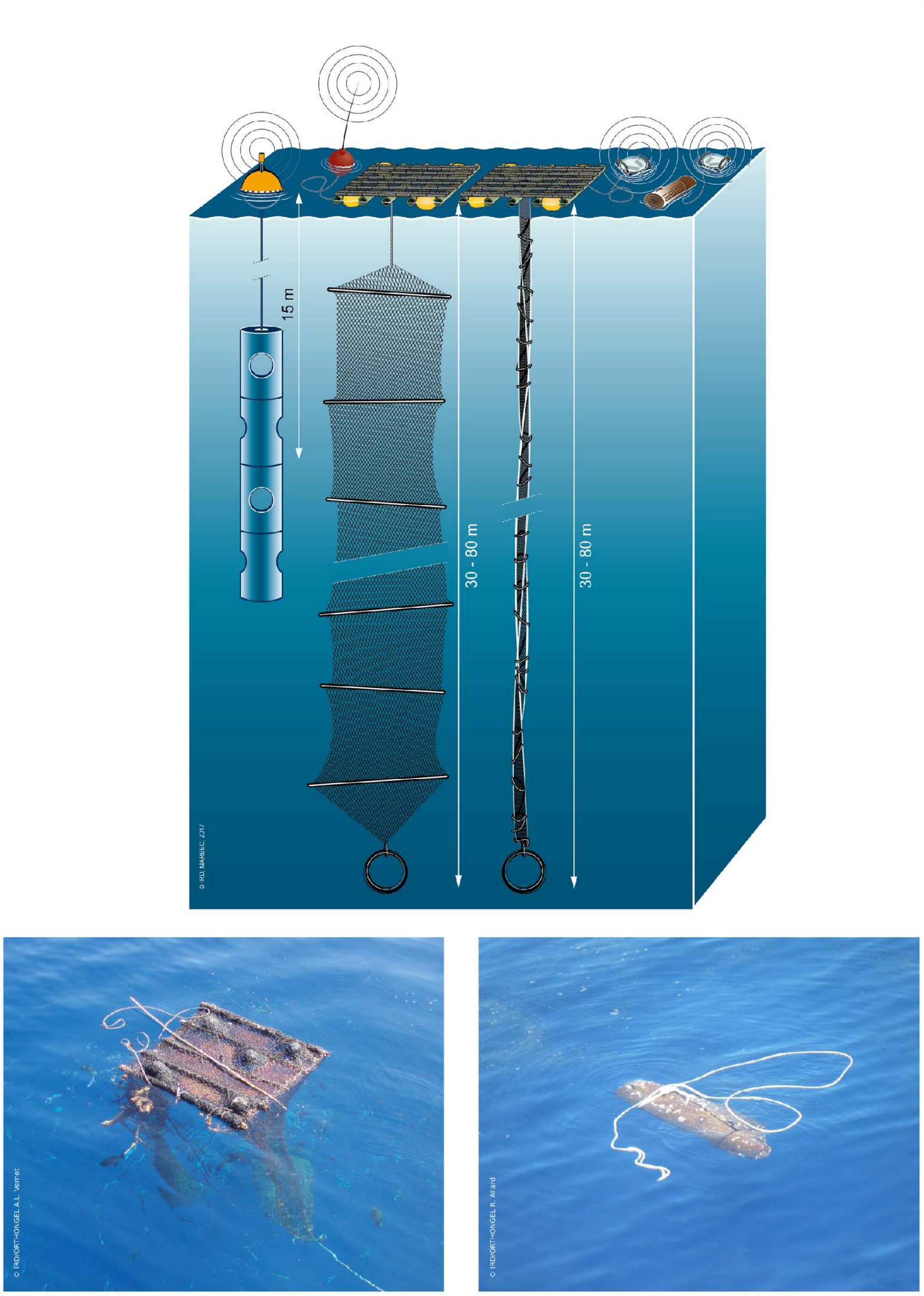

Figure 1: Description of the structure and design (in the water column) of a typical drifter (upper left) and of fish aggregating devices (FADs) used in purse seine fisheries including artificial rafts with a sea anchor made of 'curtain' nets (upper middle left, lower left photo) or 'sausage' nets (upper middle right) and natural logs (upper right, lower right photo).

*Fish Aggregating Devices*

The GPS locations of the buoys attached to the FADs used by the French fishing fleet operating in the Atlantic and Indian Oceans have been available since 2008 through a collaborative agreement between the Institut de Recherche pour le Développement (IRD) and the French frozen tuna producers' organization ORTHONGEL. The full methodology used to filter and process the raw GPS data to derive FAD trajectories at sea can be found in Maufroy et al. (2015). The current FAD data set consists of 4,777,524 positions, belonging to a total of 21,047 distinct buoys that were deployed at sea between 2008 and 2014. The sampling periodicity of FAD position varies from 15 minutes (minimum) to 2 days (maximum). It can be remotely modified to facilitate detection when a vessel is on its final approach to a FAD. Approximately 20% of the FAD data set consists of successive locations emitted within a time period of less than 6 hours and most FADs emitted two successive signals within a 24-hour period.

FADs used by the French purse seine fleets during 2008-2014 mostly consisted of rectangular bamboo rafts of about 4-6 $m^2$ covered in old pieces of purse seine nets (Franco et al., 2009). Bamboo is a light, floatable, natural composite material with a high strength-to-weight ratio that is resistant to waterlogging. Several floats made of ethylene vinyl acetate copolymer and used in the floatline of purse seine nets are generally attached under the surface structure of the raft to ensure buoyancy. In the late 2000s, a few Spanish vessels started using plastic trawl floats and PVC pipes for building FAD floating structures (Franco et al., 2009). The subsurface structure found below FADs was composed of one or two hanging panels typically made out of old purse seine netting of mesh size varying between 90 mm and 200 mm. A weight made of old pieces of chain or cables was generally attached at the bottom of the net to keep it in vertical position (Fig. 1). Initially, nets under the FAD hung in 'curtains' (Fig. 1 - upper middle left). From the early 2010s, newer design featuring 'sausages' of nets (Fig. 1 - upper middle right) were introduced to prevent accidental entanglements of turtles and sharks in the FAD's netting (ISSF, 2012; Filmalter et al., 2013). Although most of the FADs have progressively been designed with 'sausage' type nets in the Indian Ocean over years, 'curtain' type nets have remained predominant in the Atlantic Ocean. French GPS buoys have also been deployed on floating objects of natural (e.g. palm trees, logs, Fig. 1 - upper right) or anthropogenic (e.g. ropes) origins that represented about 20% of all floating objects encountered at sea by observers on French purse seiners during 2008-2014, with the Mozambique Channel being characterized by a relatively high percentage of these natural objects (Maufroy et al., 2017).

*Surface drifters*

The drifters are made up of a surface buoy ($\sim$30 cm diameter) that is attached by a long, thin tether to a holey sock drogue (sea anchor) that is centered at $\sim$15 m below the surface (Fig. 1 - upper left). The buoy measures sea surface temperature and other properties such as air pressure and wind direction and sends this information to passing satellites using an ARGOS transmitter (Lumpkin and Pazos, 2007). While the size of the buoy and drogue can vary, their drag area ratio is standardized, which acts to constrain their downwind slip (Niiler and Paduan, 1995). The GDP archives most of the data collected by the drifters. We downloaded our data set (1,092,466 positions belonging to 2,285 distinct, drogued drifters having occurred in the Indian and Atlantic Oceans during 2008-2014) from `ftp://ftp.aoml.noaa.gov/phod/pub/buoydata/`. Hansen and Poulain (1996) detail the corrections that are applied to the raw data.

*Data filtering*

A very small number of velocity values collated from the FAD database were found to be inconsistent with the maximum speed expected for ocean currents (peak speeds of 2.6 m $s^{-1}$ and 2.01 m $s^{-1}$ reported in the Agulhas Current and Gulf Stream, respectively, by Lutjeharms (2006) and Rossby (2016); maximum speed in the drifter dataset 2.9 m $s^{-1}$). We therefore removed FAD data points that had velocity values higher than the 99.99% quantile value (471.6 cm $s^{-1}$, i.e. 9.17 knots). Only 0.01% of the remaining FAD velocity values were higher than 2.9 m $s^{-1}$.

*Data distribution*

There is twice as much data for the Indian Ocean as the Atlantic Ocean but overall, the number of FAD locations has increased markedly in both oceans over the study period while the amount of drifter data remained relatively constant (Table 1). This reflects the significant expansion in the FAD fishery that has taken place in both regions (Maufroy et al., 2017). In this study, we focused on eight large biogeographical provinces, four of which occurred in the Atlantic Ocean (i.e. Guinea Current Coastal (GUIN), Eastern Tropical (ETRA), North Atlantic Tropical (NATR), and Western Tropical Atlantic (WTRA)) and four of which occurred in the Indian Ocean (East Africa Coastal (EAFR), North West Arabian Upwelling (ARAB), Indian Monsoon Gyres (MONS), and Indian Southern Subtropical Gyre (ISSG)) (Longhurst, 2007). The total number of FAD data points collated for these provinces was >50,000 (Table 2).

Table 1: Annual number of fish aggregating device (FAD) and drifter observations analyzed in the Atlantic and Indian Ocean.

| Device | ocean | 2008 | 2009 | 2010 | 2011 | 2012 | 2013 | 2014 |
|---|---|---|---|---|---|---|---|---|
| FADs | Atlantic | 17,849 | 45,469 | 102,216 | 153,990 | 286,156 | 322,490 | 464,930 |
| FADs | Indian | 105,356 | 149,211 | 200,983 | 382,315 | 580,547 | 784,130 | 1,181,882 |
| Drifters | Atlantic | 93,540 | 108,828 | 84,912 | 65,851 | 75,974 | 118,127 | 87,067 |
| Drifters | Indian | 51,482 | 38,311 | 48,314 | 46,669 | 52,378 | 80,788 | 140,225 |

*Satellite currents*

The satellite-derived surface current information produced by the OSCAR processing system is provided in near-real time from a combination of quasi-steady geostrophic and locally wind-driven dynamics (Lagerloef et al., 1999) (`http://www.oscar.noaa.gov`). The OSCAR product combines: (i) a geostrophic term computed from the gradient of ocean surface topography fields using several sources of spatial observation through time, (ii) a wind-driven velocity term computed from an Ekman-Stommel formulation with variable eddy viscosity using QuikSCAT and National Centers for Environmental Prediction winds, and (iii) a thermal wind adjustment using Reynolds sea surface temperature (Reynolds and Rayner, 2002). Dohan and Maximenko (2010) provide a full description of the OSCAR product. In this study, we used the 1/3 degree grid and 5-day interval resolution of the OSCAR currents, which is designed to represent a 30 m surface layer average. The OSCAR currents have been validated with moored buoys, drifters, and shipboard acoustic Doppler current profilers (Johnson et al., 2007).

*Direct comparison*

To compare possible velocity differences between the floating devices, we selected every FAD and drifter pair that emitted a signal in near space and time. Thus, for each FAD location and 24-hour time period, we searched for a drifter within a 1/6 degree radius (∼10 nm). If several drifters were identified, we selected the device that was closest in time. A sensitivity analysis, with time periods of 12 hours and 2.5 days (consistent with the OSCAR temporal resolution), was then conducted. The correlation between the corresponding zonal and meridional velocity components for the FAD and drifter pairs was then considered using the Pearson's correlation coefficient (Johnson et al., 2007). We then used major axis regression models (forced through the origin, based on preliminary tests indicating that estimated intercepts were generally not significantly different from 0) to assess the agreement between the two variables (Legendre and Legendre, 1998; Warton et al., 2006). This approach accounts for the measurement errors in both variables.

Table 2: Total number of fish aggregating device (FAD) and drifter observations collected in the Longhurst biogeographical provinces between 2008 and 2014. Selected provinces are shaded.

| Province description | Code | Drifters | FADs |
|---|---|---|---|
| Australia-Indonesia Coastal Province | AUSW | 16,651 | 6,106 |
| Benguela Current Coastal Province | BENG | 6,431 | 1,858 |
| Brazil Current Coastal Province | BRAZ | 19,616 | 1,399 |
| Canary Coastal Province | CNRY | 13,085 | 21,399 |
| China Sea Coastal Province | CHIN | 4,126 | 0 |
| E. Africa Coastal Province | EAFR | 31,129 | 175,733 |
| E. India Coastal Province | INDE | 10,248 | 237 |
| Guianas Coastal Province | GUIA | 12,693 | 8,946 |
| Guinea Current Coastal Province | GUIN | 8,009 | 234,069 |
| NW Arabian Upwelling Province | ARAB | 23,477 | 367,690 |
| Red Sea, Persian Gulf Province | REDS | 39 | 10 |
| Sunda-Arafura Shelves Province | SUND | 6,872 | 83 |
| SW Atlantic Shelves Province | FKLD | 189 | 0 |
| W. India Coastal Province | INDW | 5,494 | 2,401 |
| Archipelagic Deep Basins Province | ARCH | 27,079 | 554 |
| Caribbean Province | CARB | 2,515 | 174 |
| Eastern Tropical Atlantic Province | ETRA | 59,360 | 780,874 |
| Indian Monsoon Gyres Province | MONS | 146,717 | 2,665,216 |
| Indian S. Subtropical Gyre Province | ISSG | 160,585 | 162,963 |
| N. Atlantic Tropical Gyral Province | NATR | 184,484 | 63,101 |
| South Atlantic Gyral Province | SATL | 260,199 | 47,029 |
| Western Tropical Atlantic Province | WTRA | 57,414 | 230,438 |
| S. Subtropical Convergence Province | SSTC | 34,406 | 805 |
| Subantarctic Province | SANT | 1,603 | 27 |
| N. Atlantic Subtropical Gyral Province (East) | NASE | 0 | 87 |

*Indirect comparison*

This comparative analysis was then extended to the full data set by undertaking an indirect comparison of FAD and drifter velocities using satellite measurements of near-surface current velocities. At each FAD and drifter position, we linearly interpolated the OSCAR current data in time and space to calculate the OSCAR velocities (Johnson et al., 2007; Dohan and Maximenko, 2010). To determine the correlation and agreement between the FADs and OSCAR and drifters and OSCAR, we used the methodology described in the previous section. This analysis was completed at both the basin and large biogeographical province scales to ensure that the different oceanographic regimes of the Indian and Atlantic Oceans were represented. The spatio-temporal autocorrelation of velocity values along the FAD and drifter trajectories was accounted for by subsampling the data at values that were close (5 days) and far above (15 days) the Lagrangian integral time scale estimated for drifters in the Indian Ocean (i.e. 2-7 days; Peng et al., 2014).

*Projection of FAD and drifter locations using OSCAR*

The OSCAR velocities were then used to project the FAD and drifter locations from one timestep to the next to compare their Lagrangian transport in near-surface waters. We computed the distance $d$ between the projected location and the next observed location and the distance $D$ between the current location and next observed location to estimate the index $d/D$ for FADs and drifters. These indices were used to gauge the degree of departure of each floating device from the OSCAR currents predictions (Berta et al., 2014; Yaremchuk et al., 2016). Distributions were compared (i.e., for FADs and drifters) at both the basin and (selected) large biogeographical province scales.

## Results

At the basin scale, the velocity distributions of FADs and drifters were similar in the Atlantic Ocean, where the first quartile, median, and third quartile values in the FAD and drifter velocity distributions were 11.45, 19.96, 32.8 cm $s^{-1}$, and 11.59, 19.34, 30.07 cm $s^{-1}$, respectively. In the Indian Ocean, the velocity distributions of both device types were found to be different and with higher values, 21.58, 35.13, 54 cm $s^{-1}$ for FADs and 16.18, 26.39, 40.36 cm $s^{-1}$ for drifters. At a regional scale, FAD and drifter velocities were similar in the ETRA, NATR, and WTRA biogeographical provinces of the inter-tropical Atlantic Ocean, but they differed in the GUIN province (Table 3). In that province, the number of drifter locations was the lowest, more than an order of magnitude lower than the number of FAD locations (Table 2). Within the four provinces that make up most of the southwestern Indian Ocean, FAD velocities were substantially higher than drifter velocities (Table 3). Differences in velocities between FADs and drifters were attributed to differences in the spatio-temporal distribution between the two types of devices. In the Atlantic Ocean, the FAD data were concentrated in the central-eastern region (Fig. 2A) while the drifter data were more evenly distributed, although the northern area showed the highest concentrations (Fig. 2B). In the Indian Ocean, the FAD data were concentrated in the central-western region (Fig. 2A) while the drifter data were more evenly distributed over the entire basin (Fig. 2B). At a smaller, $1° \times 1°$ spatial scale, the FADs and drifters showed very similar patterns of velocity in the near-surface currents (Fig. 2C and D), revealing the major oceanographic features of both the tropical Atlantic Ocean (the South Equatorial and the North Brazil currents, the Equatorial countercurrent and the Guinea current) and the Indian Ocean (Somali, North Madagascar, and Agulhas currents, the Equatorial countercurrent and the South Equatorial current).

More than 18,000 pairs of FADs and drifters were detected across the

Table 3: The first quartile, median, and third quartile values (cm $s^{-1}$) from the fish aggregating device (FAD) and drifter velocity distributions in selected Longhurst biogeographical provinces of the Atlantic (upper part of the table) and Indian (lower part) Oceans (see Table 2 for acronyms of the provinces and Fig. 2 for their location).

| Device | Province | 1st quartile | Median | 3rd quartile |
|---|---|---|---|---|
| Drifters | ETRA | 11.62 | 19.65 | 31.45 |
| FADs | ETRA | 12.08 | 20.29 | 31.91 |
| Drifters | GUIN | 13.07 | 23.05 | 38.60 |
| FADs | GUIN | 8.91 | 15.98 | 28.49 |
| Drifters | NATR | 8.36 | 13.39 | 19.95 |
| FADs | NATR | 9.03 | 15.09 | 24.42 |
| Drifters | WTRA | 15.47 | 26.73 | 42.99 |
| FADs | WTRA | 15.58 | 27.39 | 44.18 |
| Drifters | ARAB | 14.25 | 23.86 | 39.37 |
| FADs | ARAB | 27.17 | 45.67 | 75.87 |
| Drifters | EAFR | 18.77 | 33.29 | 56.81 |
| FADs | EAFR | 22.63 | 36.72 | 55.12 |
| Drifters | ISSG | 13.70 | 22.20 | 33.39 |
| FADs | ISSG | 18.36 | 28.56 | 40.73 |
| Drifters | MONS | 17.00 | 27.89 | 43.64 |
| FADs | MONS | 21.27 | 34.51 | 52.65 |

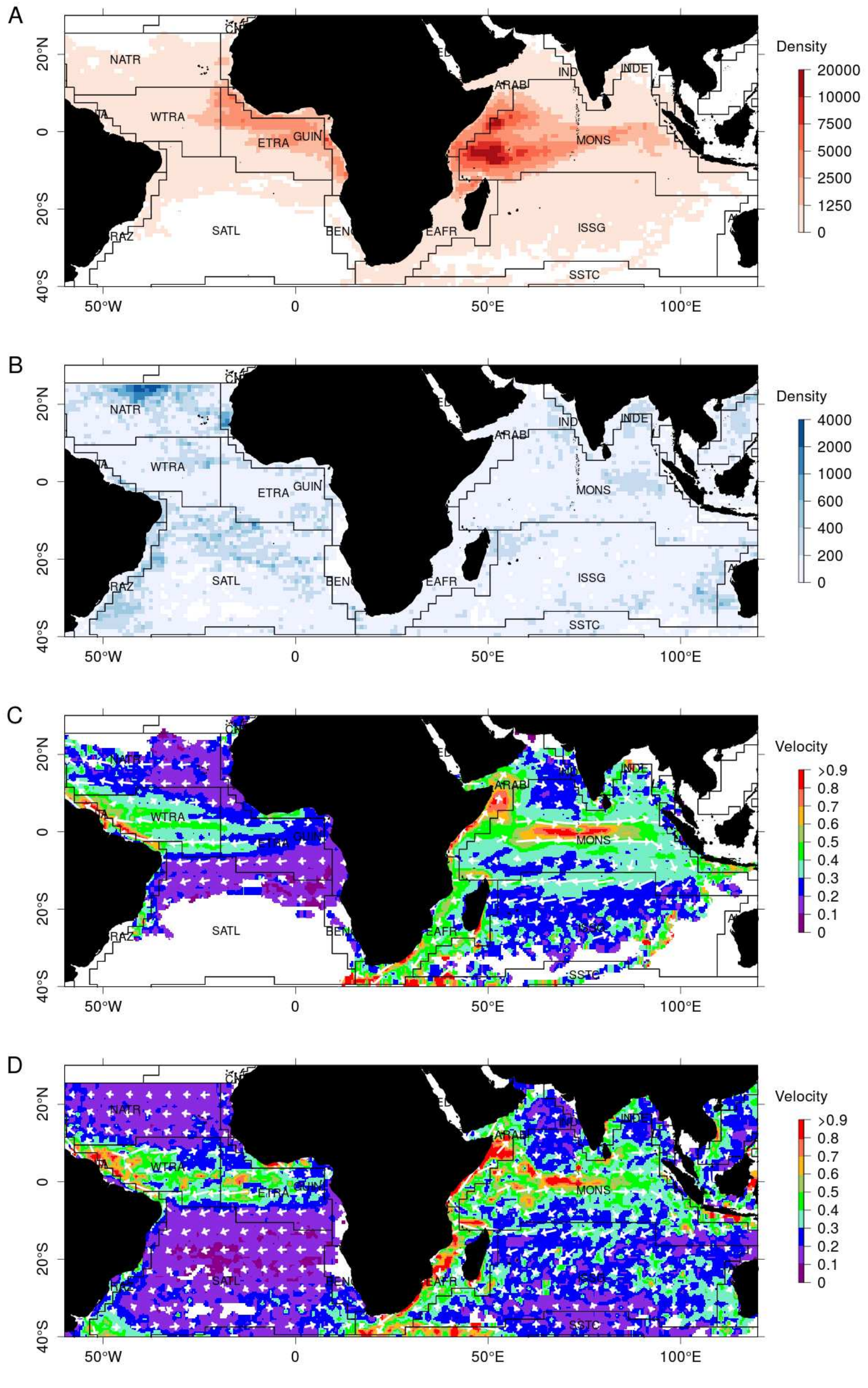

Figure 2: Spatial distribution of fish aggregating devices (FADs; A) and drifters (B) in the Atlantic and Indian Oceans. Density corresponds to the number of location points observed in each 1° × 1° grid cell for the time period 2008-2014. Mean of near-surface ocean currents (m $s^{-1}$) for the period 2008-2014, derived from FAD (C) and drifter (D) movements. Solid lines indicate boundaries between biogeographical provinces (Longhurst, 2007) (see Table 2 for acronyms).

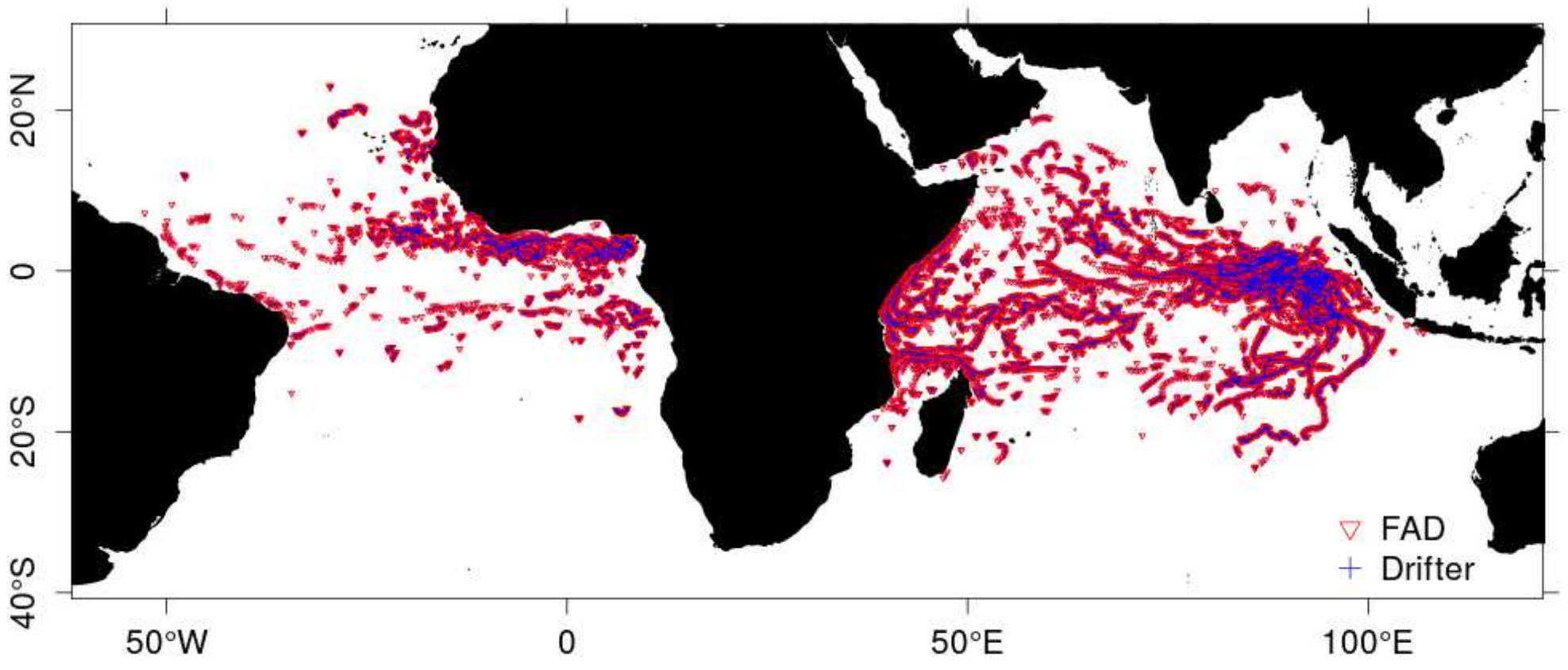

Figure 3: Spatial distribution of FADs (red triangles) and drifters (blue crosses) pairs that occurred within a 10 nm radius during 24-hour periods in the Atlantic ($n$ = 4,146) and Indian ($n$ = 14,558) Oceans.

Atlantic ($n$ = 4,146) and Indian ($n$ = 14,558) Oceans (Fig. 3). For these pairs, the zonal and meridional components of the FAD vs. drifter velocities were found to be significantly and highly correlated with Pearson's correlation coefficients between 0.68 and 0.93 (Fig. 4). This result was found to be robust to the time period considered for the definition of pairs of floating devices (Table 4). We also found several pairs in both oceans that shared common trajectories over several weeks to months, e.g., two FADs deployed in the Indian Ocean in 2013 traveled with two drifters during several months (Fig. 5). In the Indian Ocean, the velocity of FAD and drifter pairs agreed remarkably well (Fig. 4 and Table 4). In the Atlantic Ocean, however, small but consistent systematic differences in the velocity components indicate that drifters move faster than FADs (2-37% higher velocity components, 10-21% higher overall velocity; Fig. 4 and Table 4). When major axis regressions were not forced through the origin, there was no change in these results except for the velocity component in the Atlantic Ocean, for which slopes came closer to 1 and intercepts were significantly different from 0 (Appendix Table A1).

The outcomes of the comparative analysis of FAD and drifter velocities with OSCAR satellite current products further supports the case for using

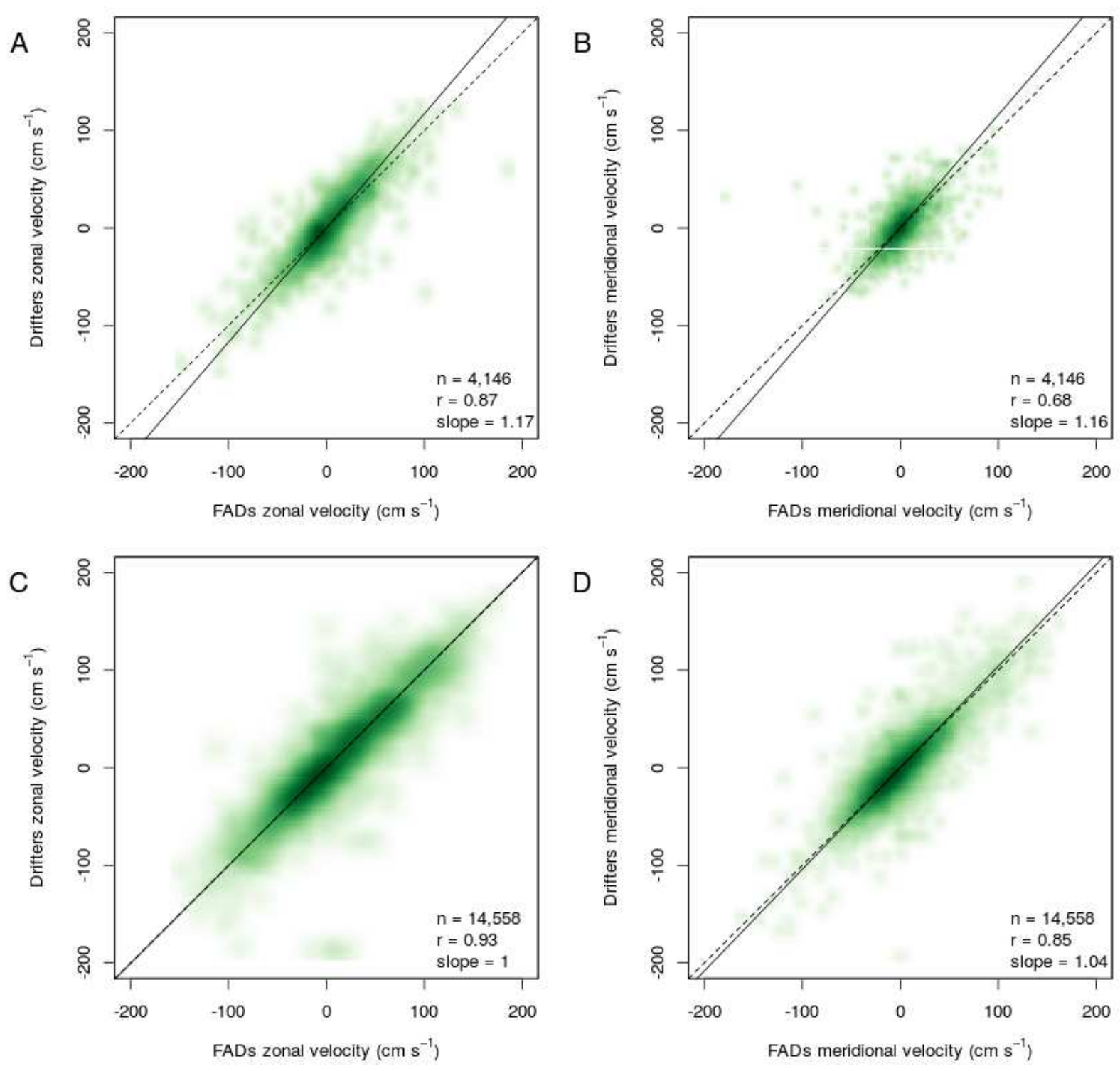

Figure 4: Velocity comparisons between the FAD and drifter pairs (A) zonal component in the Atlantic Ocean; (B) meridional component in the Atlantic Ocean; (C) zonal component in the Indian Ocean; and (D) meridional component in the Indian Ocean. The solid line indicates the major axis regression model and the dashed line indicates the 1:1 isoline.

Table 4: The number of FADs and drifters pairs, correlation coefficients and slopes of the velocity components for FAD vs. drifter at different spatio-temporal buffers in the Atlantic and Indian Oceans.

| Ocean | deltaD_deg | deltaT_day | component | n | r | Slope | Slope low. | Slope upp. |
|---|---|---|---|---|---|---|---|---|
| Indian | 1/6 | 0.5 | velocity | 10,015 | 0.85 | 1.01 | 1 | 1.01 |
| Indian | 1/6 | 1 | velocity | 14,558 | 0.83 | 1.01 | 1.01 | 1.02 |
| Indian | 1/6 | 2.5 | velocity | 25,967 | 0.78 | 1.03 | 1.02 | 1.03 |
| Atlantic | 1/6 | 0.5 | velocity | 2,842 | 0.73 | 1.12 | 1.1 | 1.14 |
| Atlantic | 1/6 | 1 | velocity | 4,146 | 0.75 | 1.15 | 1.13 | 1.16 |
| Atlantic | 1/6 | 2.5 | velocity | 7,739 | 0.71 | 1.2 | 1.18 | 1.21 |
| Indian | 1/6 | 0.5 | u | 10,015 | 0.93 | 1 | 0.99 | 1.01 |
| Indian | 1/6 | 1 | u | 14,558 | 0.93 | 1 | 1 | 1.01 |
| Indian | 1/6 | 2.5 | u | 25,967 | 0.9 | 1.01 | 1.01 | 1.02 |
| Atlantic | 1/6 | 0.5 | u | 2,842 | 0.87 | 1.16 | 1.14 | 1.19 |
| Atlantic | 1/6 | 1 | u | 4,146 | 0.87 | 1.17 | 1.15 | 1.19 |
| Atlantic | 1/6 | 2.5 | u | 7,739 | 0.85 | 1.21 | 1.19 | 1.22 |
| Indian | 1/6 | 0.5 | v | 10,015 | 0.88 | 1.02 | 1.01 | 1.03 |
| Indian | 1/6 | 1 | v | 14,558 | 0.85 | 1.04 | 1.03 | 1.05 |
| Indian | 1/6 | 2.5 | v | 25,967 | 0.76 | 1.08 | 1.07 | 1.09 |
| Atlantic | 1/6 | 0.5 | v | 2,842 | 0.69 | 1.06 | 1.02 | 1.1 |
| Atlantic | 1/6 | 1 | v | 4,146 | 0.68 | 1.16 | 1.12 | 1.2 |
| Atlantic | 1/6 | 2.5 | v | 7,739 | 0.58 | 1.33 | 1.29 | 1.37 |

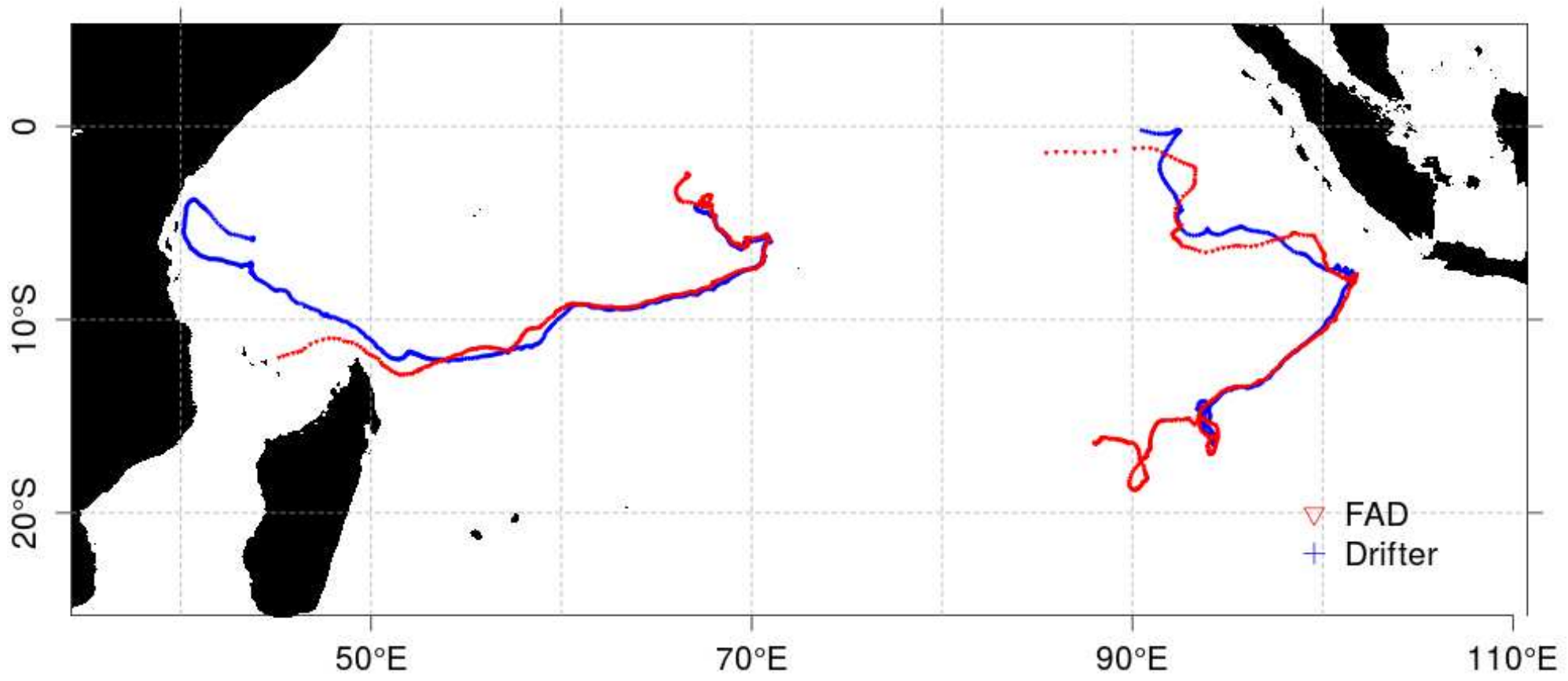

Figure 5: Examples of long-associated drift across the Indian Ocean featuring fish aggregating device (FAD) buoy n°17179 (red triangles) and drifter n°109550 (blue crosses) on the left, FAD buoy n°16812 and drifter n°109364 on the right, sharing similar trajectories between August and November 2013.

FADs for monitoring ocean surface dynamics. Here, the spatial patterns in both FAD and drifter current velocities and directions were consistent with the remotely-sensed surface currents (Appendix Fig A1). The correlation coefficients of velocity components between FADs and OSCAR and drifters and OSCAR were generally very similar (Fig. 6, Appendix Figs. A2-A5, Appendix Figs. A6-A7 and Appendix Table A2). After accounting for autocorrelation in the data, these relationships were still highly significant (Appendix Table A3). However, the OSCAR currents appeared to be slower than the currents derived from the *in-situ* data collected from the floating devices, as indicated by the slopes of the relationships between the OSCAR currents and floating devices being lower than 1 in all cases but one (Appendix Fig. A8 and Appendix Table A2). At the biogeographical province scale, the large variability observed in these slopes (FADs: 0.2-0.9 and drifters: 0.3-1.2) shows that they are not representing the surface dynamics at the same spatio-temporal scale.

The distributions of the OSCAR-projection error index $d/D$ for FADs and drifters were almost identical across all biogeographical provinces (Appendix Fig. A9), with the notable exception in the south subtropical gyre province of the Indian Ocean (ISSG). Differences in spatial coverage explain this result, with FADs mostly occurring in the North of the ISSG province during the 2008-2014 period while drifters spanned the whole area (Fig. 2).

## Discussion

We combined large data sets of remotely-sensed current speed with the GPS positions of thousands of satellite-tracked floating devices to show that the fish aggregating devices used in tuna fisheries and oceanographic drifters move similarly in near-surface ocean currents. This confirms that in tropical areas, the oceanographic information provided by satellite buoys on FADs could complement that gathered by the Global Ocean Observing System's drifter program. However, we highlighted some differences in the behaviour of FADs and drifters.

While drifters drogues are centered at 15 m below the surface, the FADs subsurface structure composed of curtain or sausage nets can go down to 50-60 m in the Indian Ocean, 80 m in the Atlantic Ocean. These differences in anchoring depth between the two types of floating devices, and between FADs, locate them in different current layers. Indeed, we noted some speed differences between the two types of floating devices, particularly in the Atlantic Ocean, which are likely related to differences in their drogue structures. In the absence of strong winds, the geostrophic balance dominates the upper ocean circulation. In this case, floating devices with different windage and drogues at different depths, or even without drogue, move at similar velocities. Conversely, higher and variable winds generate internal waves as well as Ekman currents. The former modify the mixed layer depth whereas the latter generate currents that quickly rotate with depth. In both cases, floating devices with different drogue lengths will move with different velocities and often in different directions (Poulain et al., 2009). At smaller scales, non-linear dynamics arising from wind-vorticity generate convergence and divergence regions where floating devices drogued at various depth will respond in different ways.

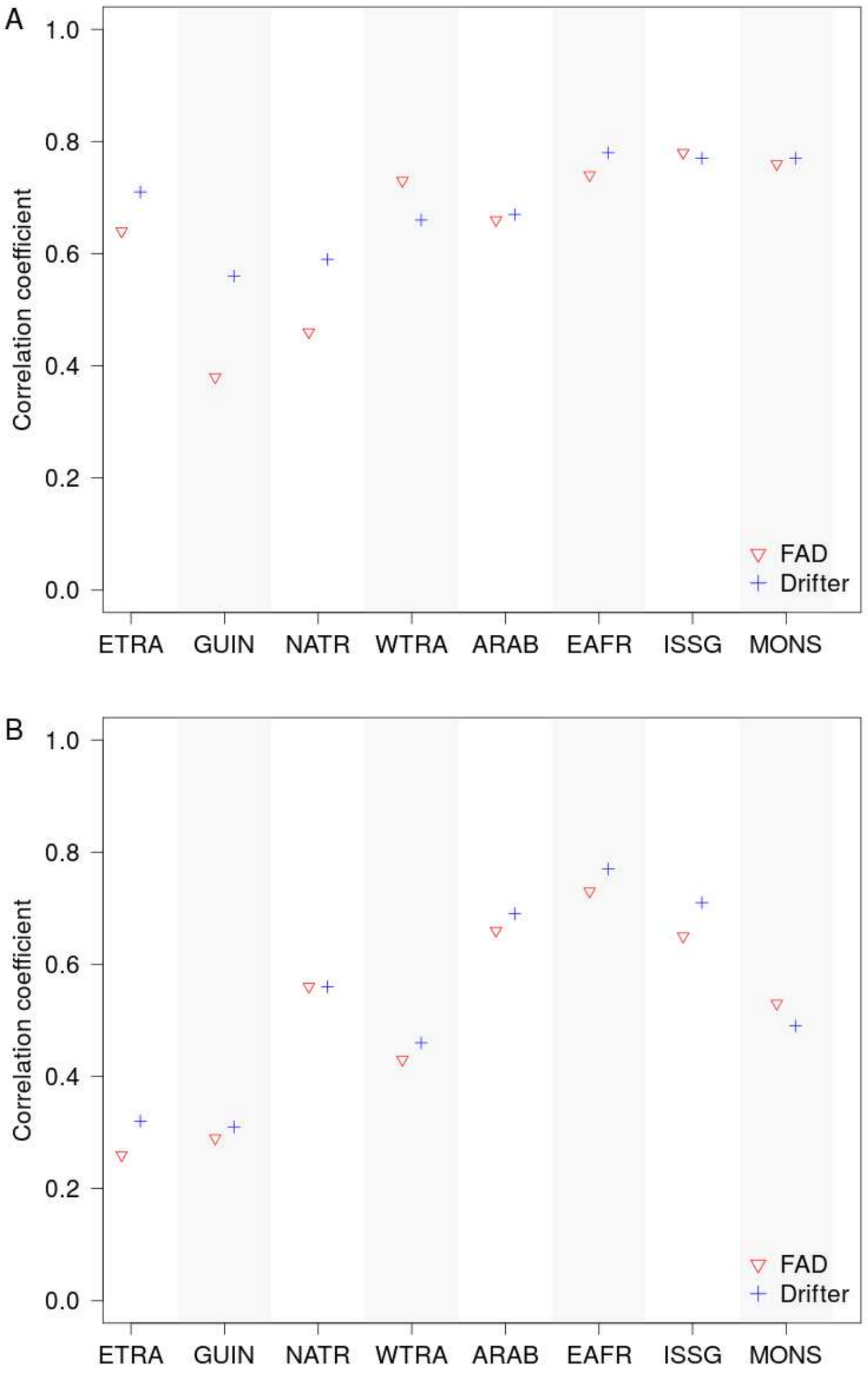

Figure 6: The comparison of correlation coefficients for the (A) zonal and (B) meridional components of velocity for the Ocean Surface Currents Analyses Real-time (OSCAR) versus fish aggregating devices (FADs) and OSCAR versus drifters in the selected Longhurst biogeographical provinces (see Table 2 for acronyms of the provinces and Fig. 2 for their location).

Velocities of drifter and FAD pairs compared remarkably well in the Indian Ocean, despite differences in their design. In that Ocean, velocities in the Equatorial countercurrent where many FADs occur have indeed been found to be relatively homogeneous along a 0-60 m depth range (Gnanaseelan and Deshpande, 2017), showing the same reversal pattern during monsoon periods. Depth homogeneous velocities were also reported along two modelled transects North of Madagascar and off Tanzania (Manyilizu et al., 2016), within two areas of high FAD occurrence. By contrast, the eastern equatorial Atlantic Ocean is characterised by the prominence of the Equatorial Undercurrent (EUC), a strong permanent eastward flow located just below the westward South Equatorial current (Johns et al., 2014). FADs built and deployed in the Atlantic Ocean have tails going down to 80 m, longer than in the other oceans (Franco et al., 2009), and at a depth where the core of the EUC is found along the equator (Johns et al., 2014). These deep tails likely slow down the drift of the FADs as compared to the shallow subsurface structure of the drifters, explaining our results.

More generally and although the mechanisms of associative behavior of tuna to FADs remain poorly understood (Fréon and Dagorn, 2000), tuna fishermen consider that deeper tails increase the attraction of tunas by slowing down the FADs (Franco et al., 2009). The depth of FAD tails were also shown to affect the tuna species composition (Lennert-Cody et al., 2008) and the arrival of fish (Orúe et al., 2017) at FADs, with tunas arriving earlier with deeper tails. Consequently, the depth of FAD appendages has been increasing in recent years in all oceans (Murua et al., 2018). In the eastern Pacific Ocean for instance, data collected by observers showed a substantial deepening of the net webbing from a median depth <10 m in the early 1990s to about 30 m nowadays (Hall and Roman, 2017). However, the progressive adoption of sausage nets in place of curtain nets, aimed at reducing the entanglement of marine species, may incidentally decrease the anchoring effect of the FAD tail appendage. In this study, our data came from fishing companies that use very similar FAD designs made of bamboo rafts and recycled fishing nets of similar lengths. More broadly, information on the structural design of FADs and their components is now being systematically collected through the fisheries observer programs run in both oceans. This new information will be useful to determine the influence of the subsurface currents on FAD drift. In particular, a comparison of FAD velocity between natural floating objects (e.g. palm trees, logs), which do not have a subsurface structure, and artificial FADs, which do have, would provide insight into the effects of design on FADs drift. A comparison of the drift and separation of concurrently deployed drifter and FAD clusters would also provide insight into the extent to which design explains the observed differences in speed between the two types of floating devices.

Given that the FAD data we used in this study is open access, we expect that further analysis will be undertaken to fully validate the potential applications of FAD data for oceanographers, and that the results of this work will prompt long-term collaborations with the tuna fishing industry. The quantity of information available to the scientific community would strongly benefit from the release of data from other purse seine fishing companies operating in the Atlantic and Indian Oceans since the French purse seine fleet only represented about 10% of the total drifting FADs in recent years (Maufroy et al., 2017). Recent availability of FAD GPS positions in the western and central Pacific Ocean shows a positive step in this direction (Escalle et al., 2017). It would also be beneficial to apply the GDP's quality control procedures (Hansen and Poulain, 1996; Lumpkin and Pazos, 2007) to the FAD data. This step may provide useful information that is currently missing such as FAD location errors.

The GPS buoys tracked in the present study were mostly deployed within the fishing grounds of the French purse seine fleet (Maufroy et al., 2015; Snouck-Hurgronje et al., 2018). Other purse seine fleets include some non-fishing support vessels that maintain the array of FADs and can deploy buoys outside fishing grounds, anticipating their drift in productive areas several weeks in advance (Arrizabalaga et al., 2001; Assan et al., 2015). In this context, GPS buoy data from fleets assisted by support vessels would greatly complement the French data set and provide a more complete picture of the near-surface ocean currents of the tropical areas covered in the present study.

More broadly, the conspicuous character of global changes presents some serious observational challenges. Effectively responding to these challenges requires better integration across individual networks and multiple platforms, to make the most of synergies between the different types of ocean observations (Roemmich et al., 2010). The development of standards for metadata and data formats, as well as access protocols (e.g., Web Services), has recently enhanced interoperability functions in information systems. Thus, these standards are better able to merge and process heterogeneous data sets stored in distributed infrastructures and promote integration across scientific disciplines (Reichman et al., 2011; Mooney et al., 2013; Robertson et al., 2014). Data management systems should also include well-described control procedures that aim to inform users about the best quality data sets available (Roemmich et al., 2010). In oceanography, the recent introduction of key standards contributes to this higher level of interoperability for physical and chemical parameters delivered as gridded data (e.g. model outputs, or satellite remote-sensing products) or time series of parameters retrieved from platforms at sea (Hankin et al., 2010). Like the data collected through citizen science initiatives (Lauro et al., 2014), the millions of data collected by fishermen could substantially increase the spatio-temporal coverage of ocean observations in a cost-efficient manner. Thus, the major contributions these data sets could potentially make to the GOOS and GCOS calls for improved collaboration with the fishing industry (Gawarkiewicz and Malek Mercer, 2018; Moreno et al., 2016) and the establishment of a system that adequately acknowledges the contributors and fosters a data sharing environment.

The openness of anonymized FAD tracking data has almost no cost for the fishing industry and provides an ideal opportunity to communicate in a transparent way about their practices. In particular, it shows willingness with regards to compliance and accountability on the limited number of active buoys per fishing vessel recently implemented by most tuna Regional Fisheries Management Organisations (RFMOs). Complying with Conservations and Management Measures of the RMFOs will increasingly become important for the allocation of stock quotas and access rights in the future (IOTC, 2018). In the Seychelles, access to some fishing grounds of the exclusive economic zone identified as part of the ongoing Management Spatial Planning will be restricted to sustainable fishing practices, which could include availability of FAD data for scientific and monitoring purpose. Globally, most purse seine fishing companies are now involved in Fisheries Improvement Projects with the objective of reaching the standards of the Marine Stewardship Certification (MSC) and increase benefits. The provision of information on FAD-related fishing practices is a key component of MSC assessment due to the growing concerns of FAD fishing (Fonteneau et al., 2013; Davies et al., 2017). Fishermen who voluntarily release data sets that are useful for advancing our understanding of ocean dynamics will benefit from their efforts through improved image and communication to the general public.

**Acknowledgements**

We are grateful to ORTHONGEL and the fishing companies CFTO, SAPMER, and SAUPIQUET for making their FAD buoy data available and open access. We are particularly grateful to Laurent Pinault, Sarah Le Couls, Anthony Claude, Martin Denniel, Gildas Bodilis, David Kaplan and Laurent Floch for their assistance with the data management and interpretation. Many thanks to Iñigo Krug for sharing information on FAD design and Pierre Lopez for his assistance in preparing the artwork. We would also like to thank the AOML's drifting buoy group for making their drifter data available and acknowledge the OSCAR Project Office for providing the surface current data. Thanks to Pierrick Penven for pointing out some useful references, Daniel Gaertner for his statistical advice, and Joël Sudre, Thierry Delcroix, Eric Greiner, Nathalie Bodin, Rick Lumpkin and three anonymous reviewers for their useful comments on earlier versions of this study. This work received funding from France Filière Pêche (FFP) through the PhD. grant of AM (FFP PH/2012/14 and IRD 303291/00), the European project CECOFAD (MARE/2012/24), the European Union's Horizon 2020 research and innovation program under grant agreements No. 675680 (BlueBridge project) and No. 731011 (OpenAIRE Connect project), and the Observa-toire Thonier of IRD. We are grateful to Jane Alpine and Anne-Elise Nieblas for carefully editing the manuscript.

## Appendix

Table A1: Same as Table 4 but with major axis regression not forced through the origin (the estimated intercept is added in the last column).

| Ocean | deltaD_deg | deltaT_day | component | n | r | Slope | Slope low. | Slope upp. | Intercept |
|---|---|---|---|---|---|---|---|---|---|
| Indian | 1/6 | 0.5 | velocity | 10,015 | 0.85 | 1.01 | 0.99 | 1.02 | 0 |
| Indian | 1/6 | 1 | velocity | 14,558 | 0.83 | 1.01 | 1 | 1.02 | 0.37 |
| Indian | 1/6 | 2.5 | velocity | 25,967 | 0.78 | 1 | 0.99 | 1.01 | 1.55 |
| Atlantic | 1/6 | 0.5 | velocity | 2,842 | 0.73 | 1.01 | 0.98 | 1.05 | 3.62 |
| Atlantic | 1/6 | 1 | velocity | 4,146 | 0.75 | 1.04 | 1.01 | 1.06 | 4.12 |
| Atlantic | 1/6 | 2.5 | velocity | 7,739 | 0.71 | 1.09 | 1.06 | 1.11 | 4.05 |
| Indian | 1/6 | 0.5 | u | 10,015 | 0.93 | 1 | 0.99 | 1.01 | -0.03 |
| Indian | 1/6 | 1 | u | 14,558 | 0.93 | 1 | 1 | 1.01 | -0.22 |
| Indian | 1/6 | 2.5 | u | 25,967 | 0.9 | 1.02 | 1.01 | 1.02 | -0.22 |
| Atlantic | 1/6 | 0.5 | u | 2,842 | 0.87 | 1.16 | 1.14 | 1.19 | 0 |
| Atlantic | 1/6 | 1 | u | 4,146 | 0.87 | 1.17 | 1.15 | 1.19 | 0.07 |
| Atlantic | 1/6 | 2.5 | u | 7,739 | 0.85 | 1.2 | 1.19 | 1.22 | 0.76 |
| Indian | 1/6 | 0.5 | v | 10,015 | 0.88 | 1.02 | 1.01 | 1.03 | -0.28 |
| Indian | 1/6 | 1 | v | 14,558 | 0.85 | 1.04 | 1.03 | 1.05 | -0.47 |
| Indian | 1/6 | 2.5 | v | 25,967 | 0.76 | 1.08 | 1.07 | 1.09 | -0.67 |
| Atlantic | 1/6 | 0.5 | v | 2,842 | 0.69 | 1.06 | 1.02 | 1.1 | -0.45 |
| Atlantic | 1/6 | 1 | v | 4,146 | 0.68 | 1.16 | 1.12 | 1.2 | -0.37 |
| Atlantic | 1/6 | 2.5 | v | 7,739 | 0.58 | 1.33 | 1.29 | 1.37 | -0.61 |

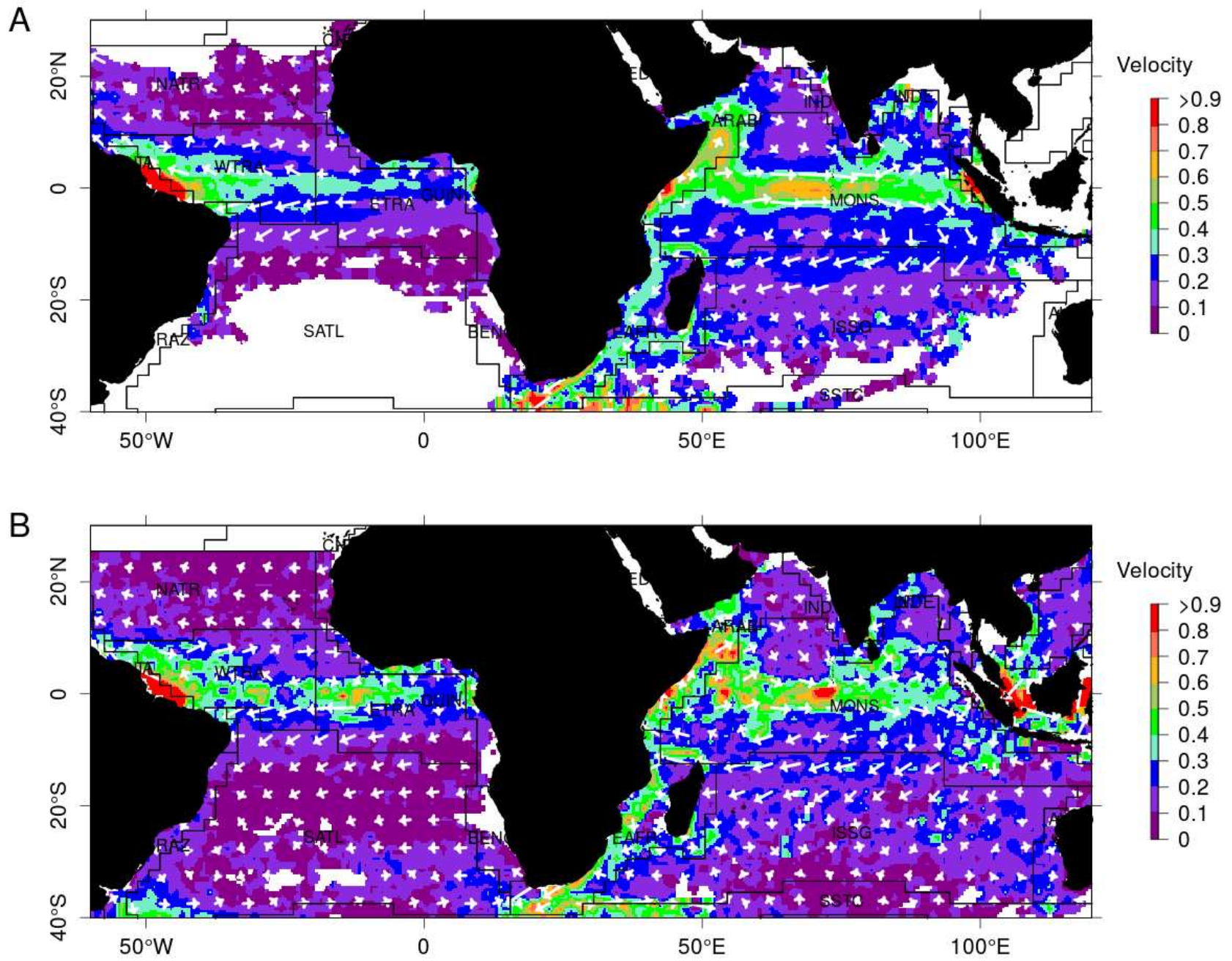

Figure A1: Mean of near-surface ocean currents (m $s^{-1}$) for the period 2008-2014, derived from OSCAR at FADs (i.e., OSCAR data interpolated at the time and location of FAD data) (A) and OSCAR at drifters (i.e., OSCAR data interpolated at the time and location of drifter data) (B). Solid lines indicate boundaries between biogeographical provinces (Longhurst, 2007) (see Table 2 for acronyms).

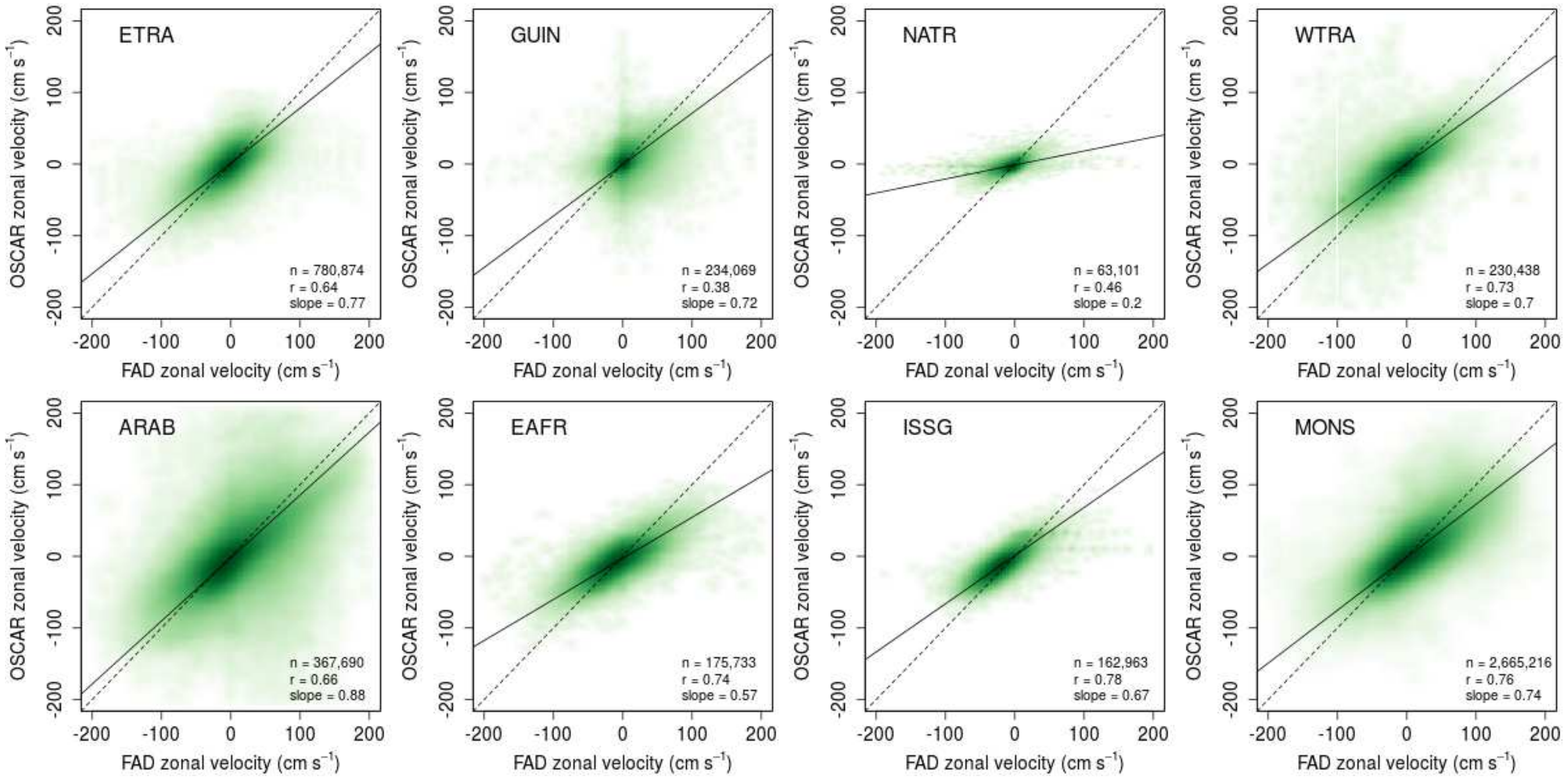

Figure A2: The comparison of zonal velocities between fish aggregating devices (FADs) and Ocean Surface Currents Analyses Real-time (OSCAR) in the selected Longhurst biogeographical provinces of the Atlantic Ocean (top) and Indian Ocean (bottom). The solid line indicates the major axis regression model and the dashed line indicates the 1:1 isoline.

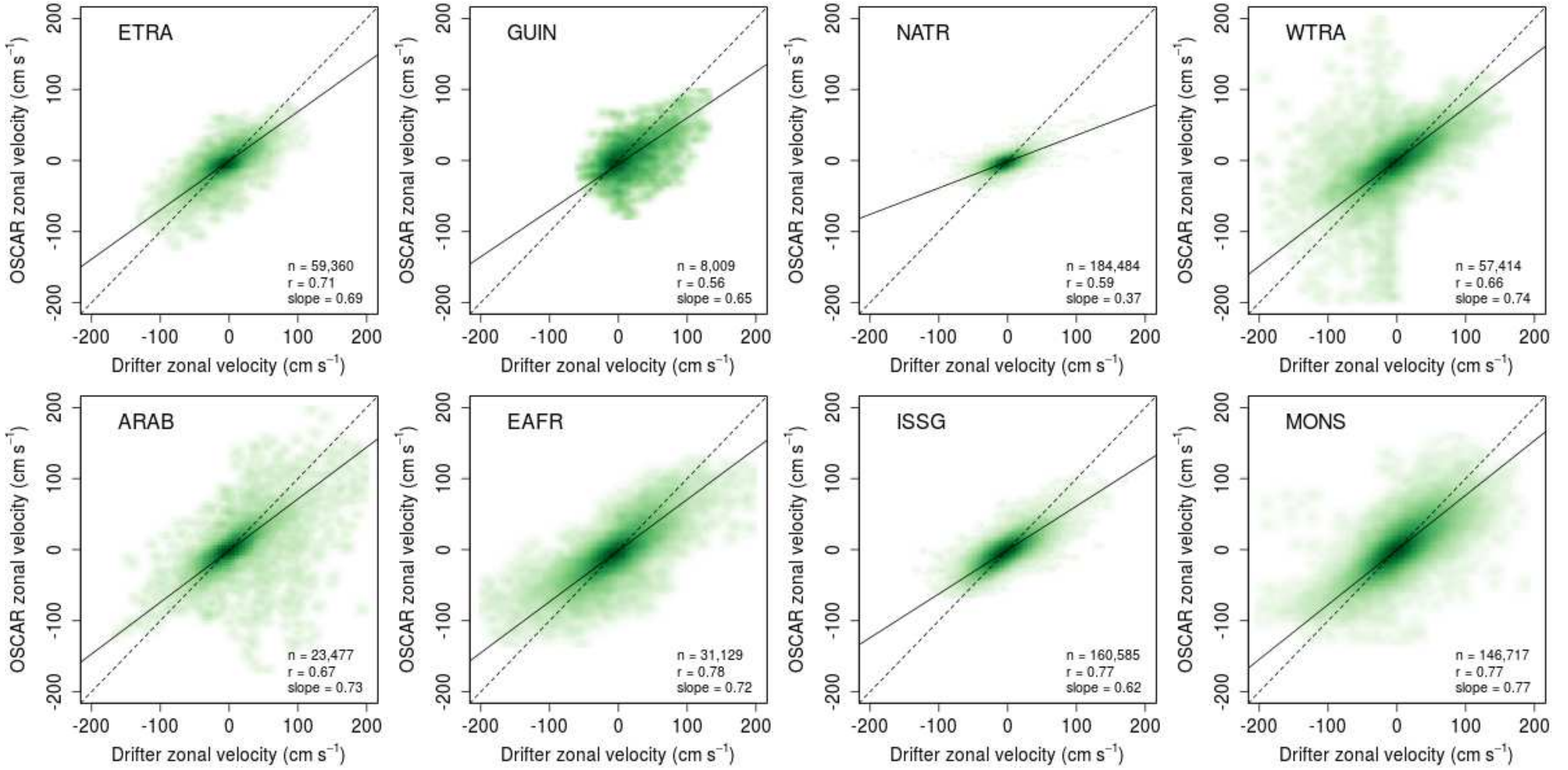

Figure A3: The comparison of zonal velocities between drifters and Ocean Surface Currents Analyses Real-time (OSCAR) in the selected Longhurst biogeographical provinces of the Atlantic Ocean (top) and Indian Ocean (bottom). The solid line indicates the major axis regression model and the dashed line indicates the 1:1 isoline.

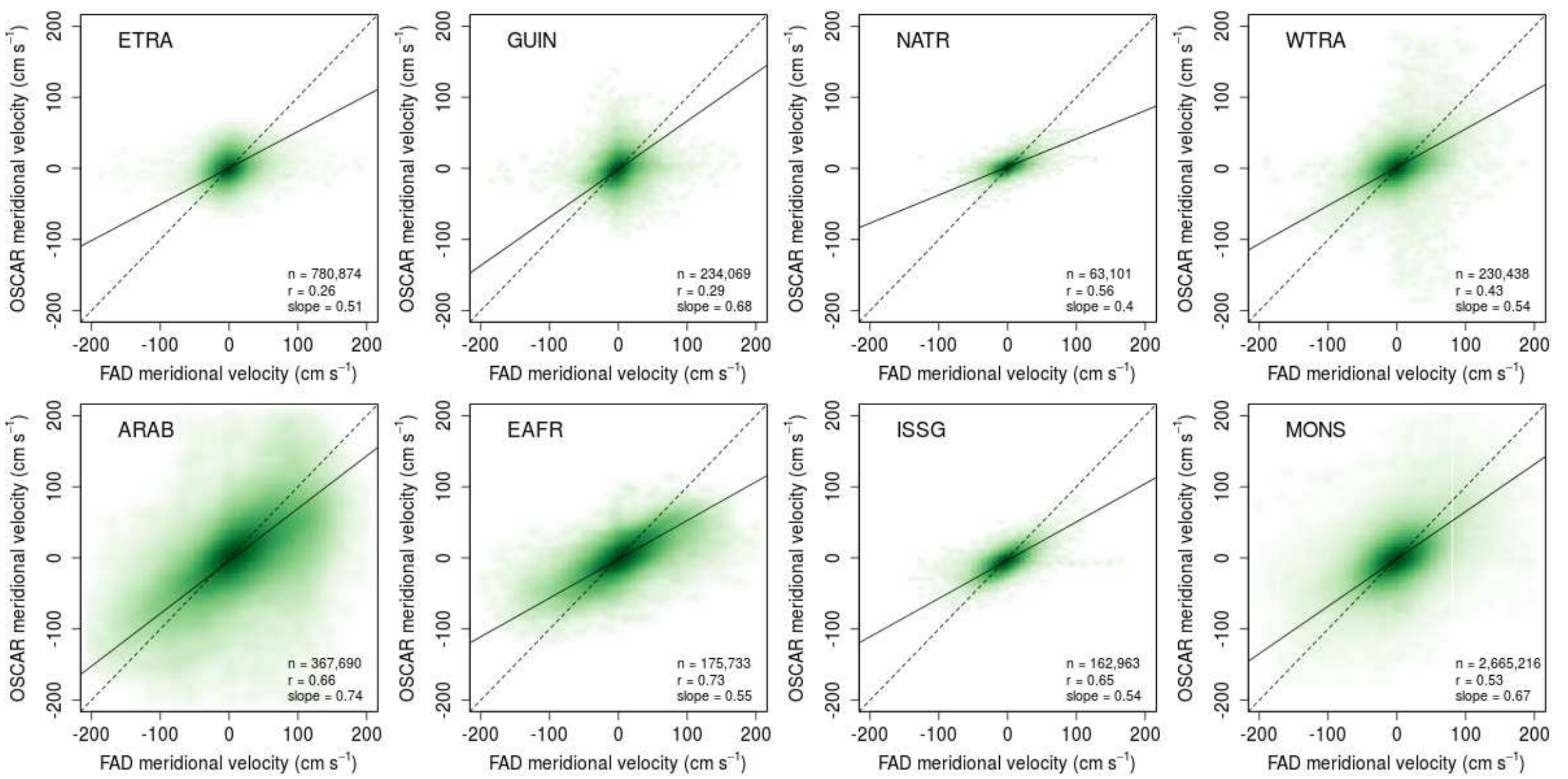

Figure A4: The comparison of meridional velocities between fish aggregating devices (FADs) and Ocean Surface Currents Analyses Real-time (OSCAR) in the selected Longhurst biogeographical provinces of the Atlantic Ocean (top) and Indian Ocean (bottom). The solid line indicates the major axis regression model and the dashed line indicates the 1:1 isoline.

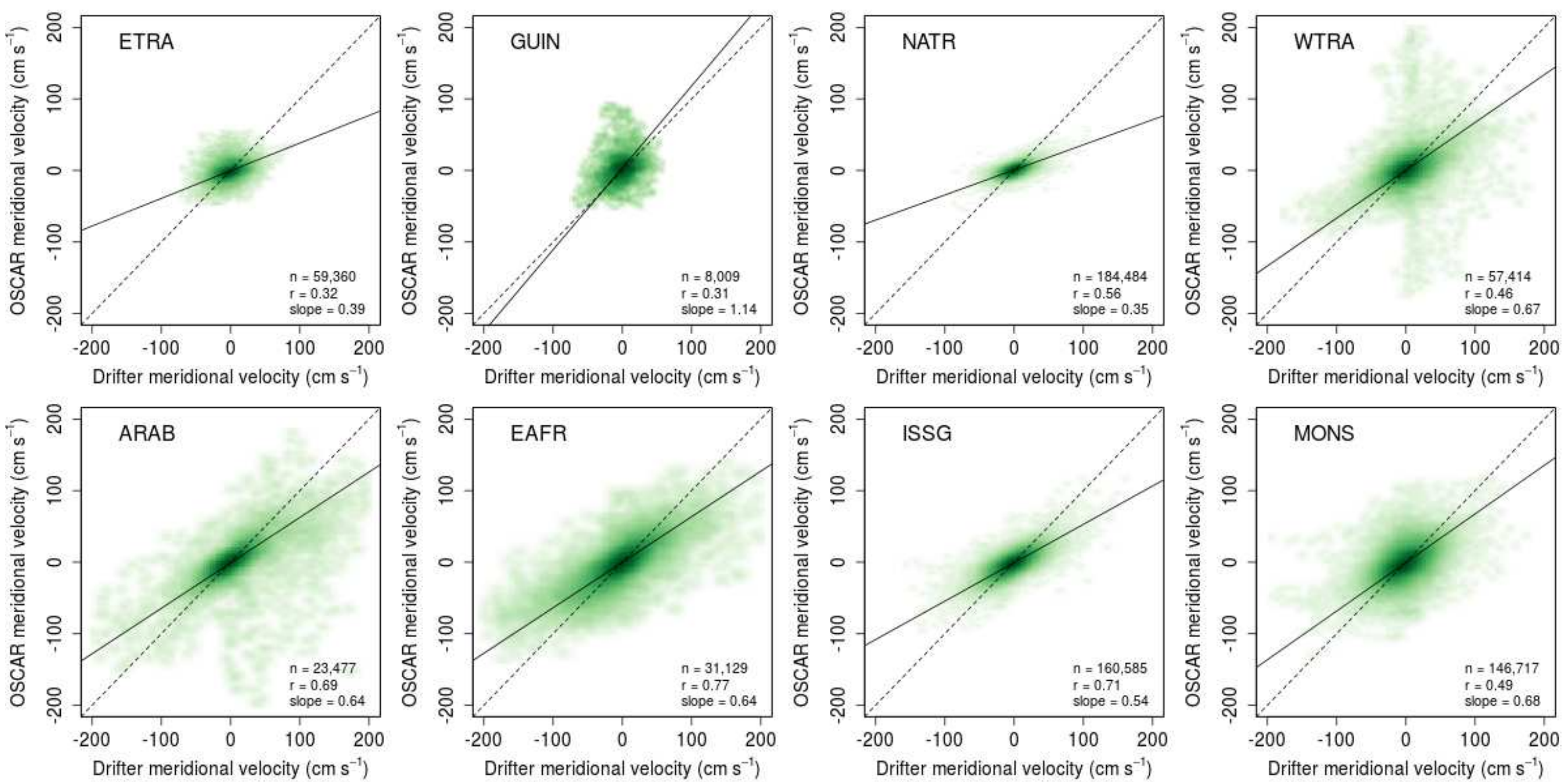

Figure A5: The comparison of meridional velocities between drifters and Ocean Surface Currents Analyses Real-time (OSCAR) in the selected Longhurst biogeographical provinces of the Atlantic Ocean (top) and Indian Ocean (bottom). The solid line indicates the major axis regression model and the dashed line indicates the 1:1 isoline.

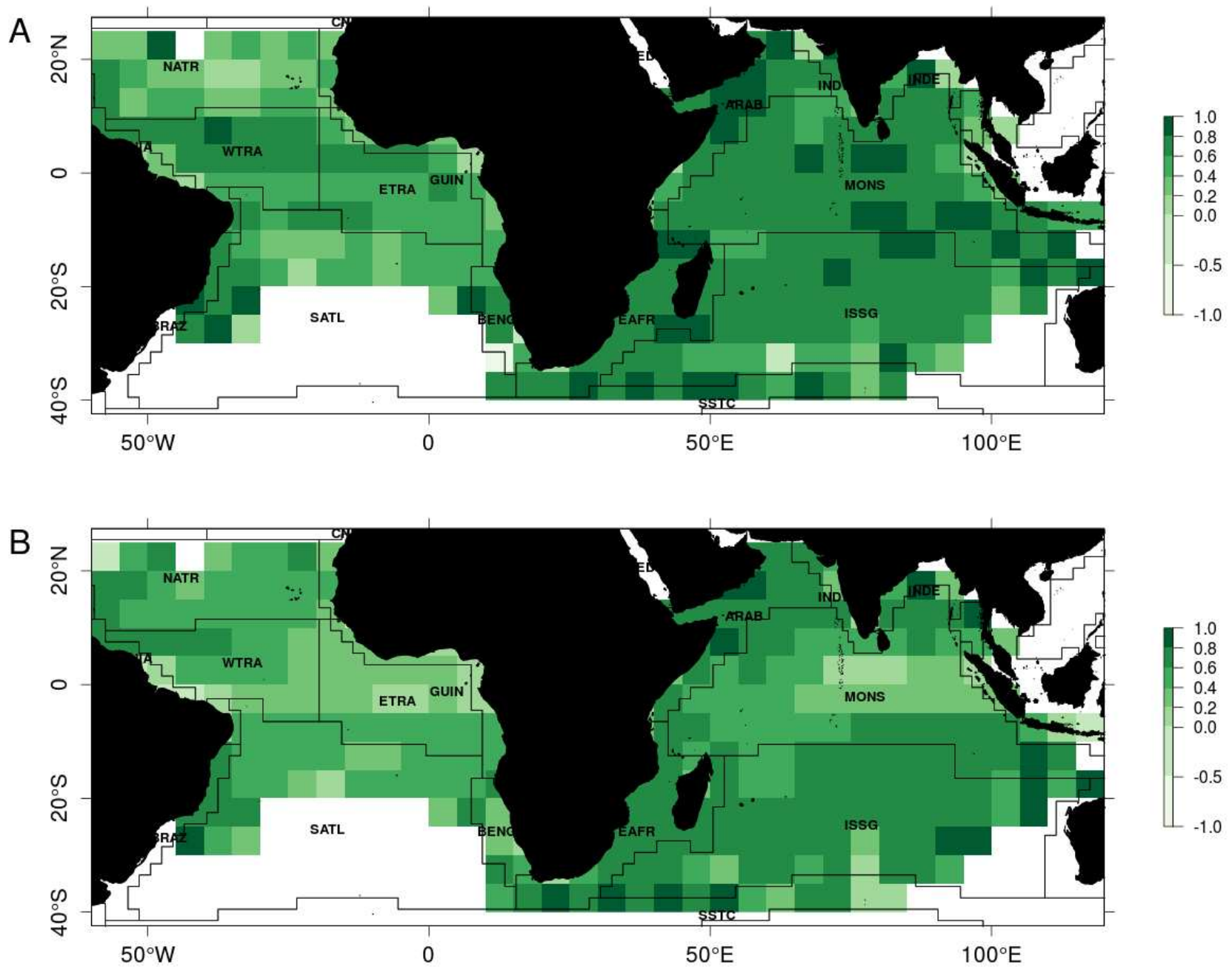

Figure A6: Correlation in each $5^\circ \times 5^\circ$ grid cell of zonal (A) and meridional (B) velocity components between FADs and OSCAR.

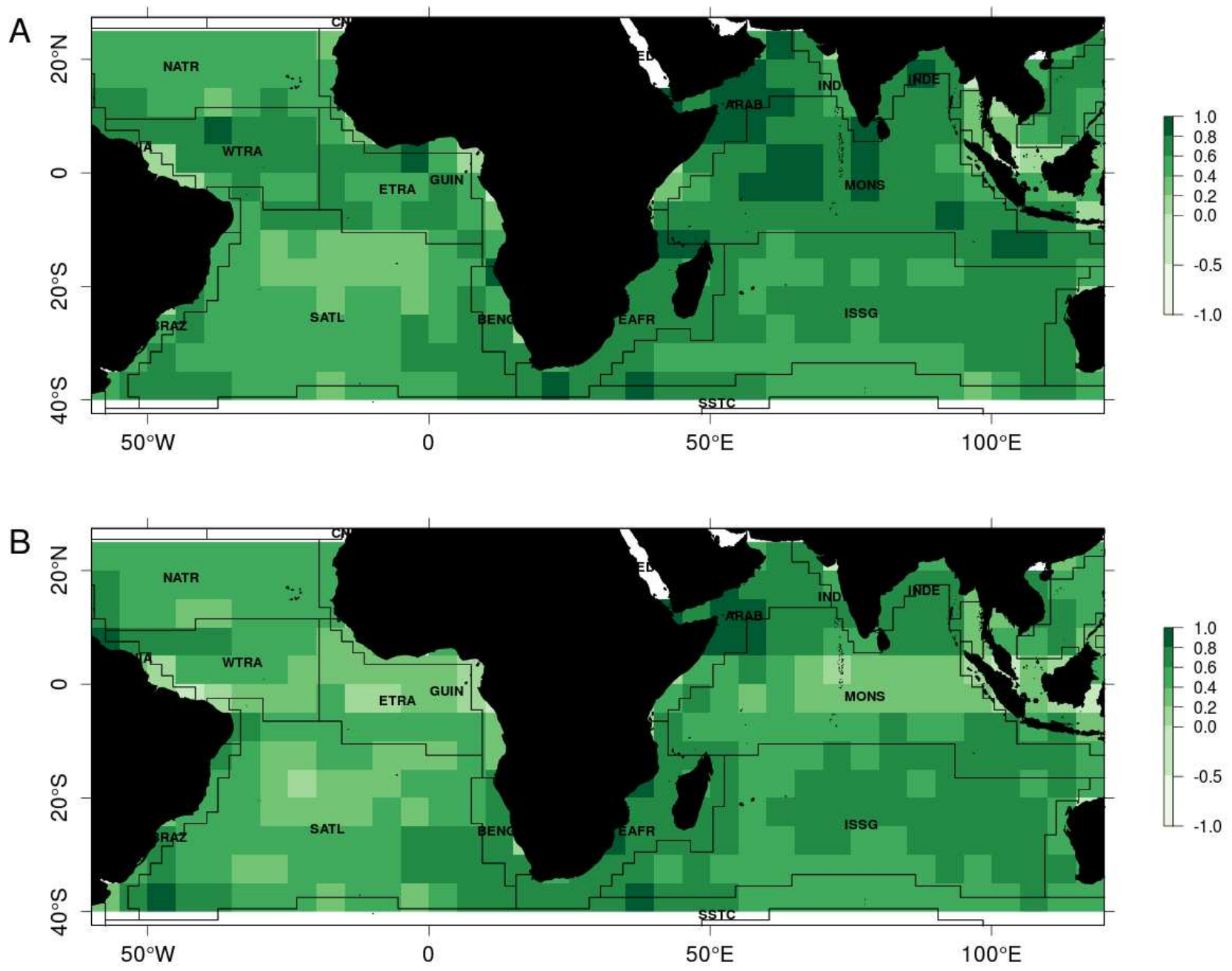

Figure A7: Correlation in each 5° × 5° grid cell of zonal (A) and meridional (B) velocity components between drifters and OSCAR.

Table A2: Summary of the major axis regression models fitted to the velocity components of the Ocean Surface Currents Analyses Real-time (OSCAR) measurements versus fish aggregating devices (FADs) and OSCAR versus drifters in the selected Longhurst biogeographical provinces of the Atlantic and Indian Oceans. Slope low. = 2.5% quantile value used as the lower limit of the regression slope estimate; Slope upp. = 97.5% quantile value used as the upper limit of the regression slope estimate. r = Pearson's correlation coefficient.

| Device | Component | Province | Slope low. | Slope | Slope upp. | r |
|---|---|---|---|---|---|---|
| FADs | Zonal | ETRA | 0.769 | 0.771 | 0.773 | 0.64 |
| FADs | Zonal | GUIN | 0.710 | 0.717 | 0.725 | 0.38 |
| FADs | Zonal | NATR | 0.192 | 0.195 | 0.198 | 0.46 |
| FADs | Zonal | WTRA | 0.696 | 0.699 | 0.702 | 0.73 |
| FADs | Zonal | ARAB | 0.876 | 0.879 | 0.882 | 0.66 |
| FADs | Zonal | EAFR | 0.571 | 0.573 | 0.576 | 0.74 |
| FADs | Zonal | ISSG | 0.669 | 0.672 | 0.675 | 0.78 |
| FADs | Zonal | MONS | 0.739 | 0.739 | 0.740 | 0.76 |
| Drifters | Zonal | ETRA | 0.686 | 0.692 | 0.697 | 0.71 |
| Drifters | Zonal | GUIN | 0.630 | 0.651 | 0.673 | 0.56 |
| Drifters | Zonal | NATR | 0.369 | 0.371 | 0.373 | 0.59 |
| Drifters | Zonal | WTRA | 0.737 | 0.744 | 0.751 | 0.66 |
| Drifters | Zonal | ARAB | 0.717 | 0.728 | 0.738 | 0.67 |
| Drifters | Zonal | EAFR | 0.713 | 0.719 | 0.725 | 0.78 |
| Drifters | Zonal | ISSG | 0.615 | 0.617 | 0.620 | 0.77 |
| Drifters | Zonal | MONS | 0.767 | 0.770 | 0.773 | 0.77 |
| FADs | Meridional | ETRA | 0.506 | 0.511 | 0.515 | 0.26 |
| FADs | Meridional | GUIN | 0.668 | 0.677 | 0.686 | 0.29 |
| FADs | Meridional | NATR | 0.391 | 0.395 | 0.400 | 0.56 |
| FADs | Meridional | WTRA | 0.534 | 0.538 | 0.543 | 0.43 |
| FADs | Meridional | ARAB | 0.736 | 0.739 | 0.742 | 0.66 |
| FADs | Meridional | EAFR | 0.543 | 0.545 | 0.547 | 0.73 |
| FADs | Meridional | ISSG | 0.535 | 0.538 | 0.541 | 0.65 |
| FADs | Meridional | MONS | 0.664 | 0.666 | 0.667 | 0.53 |
| Drifters | Meridional | ETRA | 0.377 | 0.387 | 0.396 | 0.32 |
| Drifters | Meridional | GUIN | 1.068 | 1.144 | 1.227 | 0.31 |
| Drifters | Meridional | NATR | 0.349 | 0.351 | 0.353 | 0.56 |
| Drifters | Meridional | WTRA | 0.658 | 0.669 | 0.679 | 0.46 |
| Drifters | Meridional | ARAB | 0.627 | 0.635 | 0.644 | 0.69 |
| Drifters | Meridional | EAFR | 0.631 | 0.636 | 0.642 | 0.77 |
| Drifters | Meridional | ISSG | 0.534 | 0.537 | 0.539 | 0.71 |
| Drifters | Meridional | MONS | 0.673 | 0.680 | 0.686 | 0.49 |

Table A3: The number of observations and correlation coefficients of the velocity components of Ocean Surface Currents Analyses Real-time (OSCAR) versus fish aggregating devices (FADs) and OSCAR versus drifters for the entire dataset (n, Corr_u, Corr_v) and for the datasets subsampled every 5 days (n5, corr5_u, corr5_v) and 15 days (n15, corr15_u, corr15_v) in the Atlantic and Indian Oceans.

| Device | Ocean | n | Corr_u | Corr_v | n5 | corr5_u | corr5_v | n15 | corr15_u | corr15_v |
|---|---|---|---|---|---|---|---|---|---|---|
| FADs | Atlantic | 1,393,100 | 0.62 | 0.31 | 66,228 | 0.57 | 0.27 | 36,986 | 0.53 | 0.24 |
| FADs | Indian | 3,384,424 | 0.75 | 0.58 | 181,193 | 0.72 | 0.56 | 108,611 | 0.70 | 0.55 |
| Drifters | Atlantic | 634,297 | 0.58 | 0.49 | 32,303 | 0.58 | 0.47 | 11,211 | 0.57 | 0.49 |
| Drifters | Indian | 458,065 | 0.74 | 0.63 | 23,418 | 0.75 | 0.64 | 8,187 | 0.77 | 0.65 |

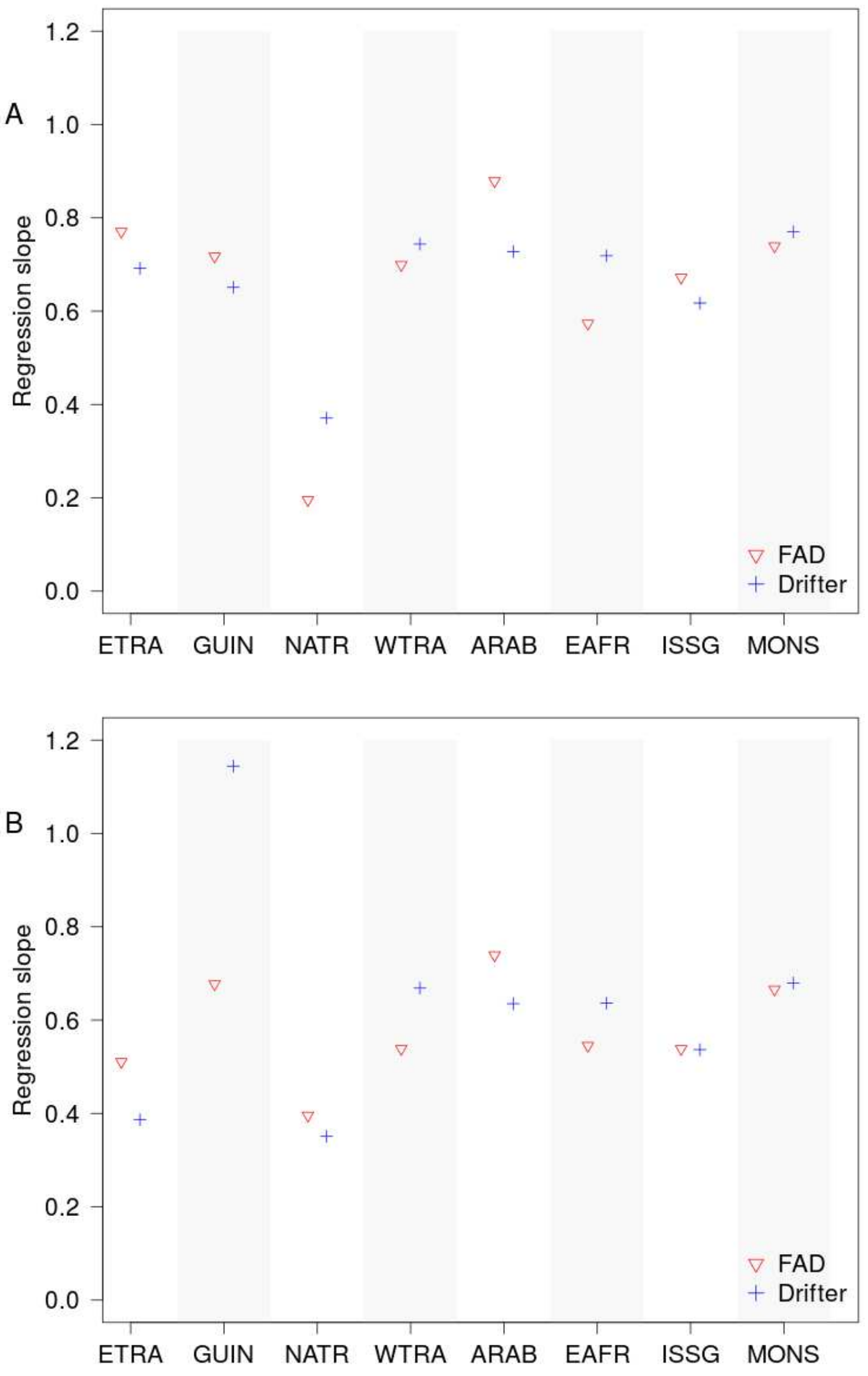

Figure A8: The comparison of slopes of major axis regression models fitted to the (A) zonal and (B) meridional velocity data of the Ocean Surface Currents Analyses Real-time (OSCAR) versus fish aggregating devices (FADs) and OSCAR versus drifters in the selected Longhurst biogeographical provinces.

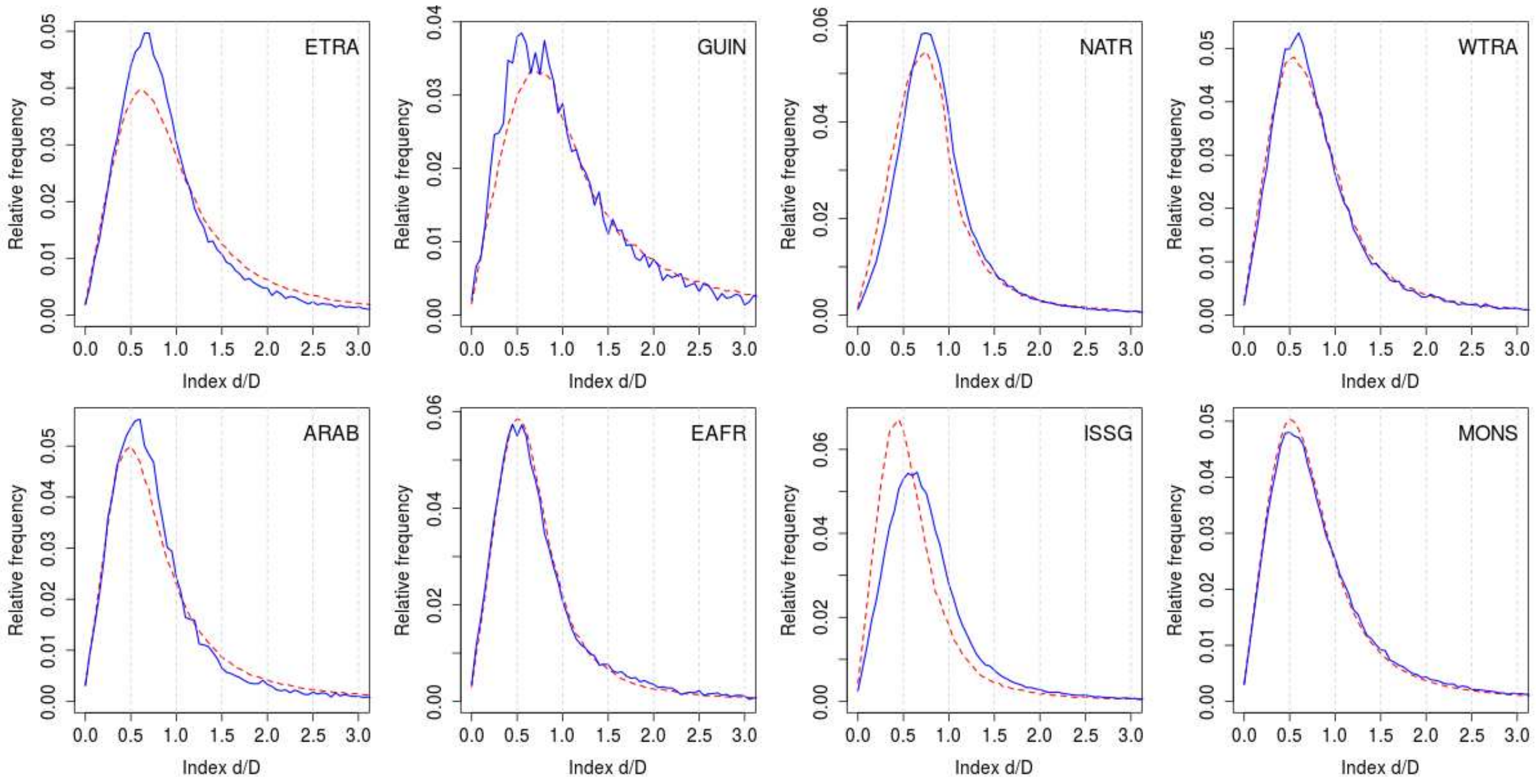

Figure A9: Relative frequency distributions of the index $d/D$ that describe the prediction skill of fish aggregating devices (FADs; dashed red curves) and drifters (solid blue curves), with respect to the Ocean Surface Currents Analyses Real-time (OSCAR) velocities for the selected Longhurst biogeographical provinces, where $d$ is the distance between the projected and observed location at the next time step and $D$ is the distance between the current and next observed locations.